\documentclass[ reprint,
amsmath,amssymb,
aps,
]{revtex4-2}

\usepackage{epsf,epsfig}
\usepackage{amssymb,amsmath,amsfonts}
\usepackage{color}
\usepackage{braket}
\usepackage{subcaption}
\usepackage[T1]{fontenc} 

\usepackage[dvipsnames, svgnames,  x11names ]{xcolor}
\usepackage[colorlinks]{hyperref}
\hypersetup{
   citecolor = {blue},
   linkcolor = {RubineRed},
}

\usepackage{physics}
\usepackage{tikz}
\usetikzlibrary{arrows,snakes,backgrounds}

\newcommand{\ra}{\rangle}
\newcommand{\be}{\begin{equation}}
\newcommand{\ee}{\end{equation}}
\newcommand{\CK}{\mathcal{K}}

\newcommand{\CC}{\mathcal{C}}
\newcommand{\CH}{\mathcal{H}}

\usepackage[english]{babel}

\usepackage{graphicx}

\definecolor{NordCyan}{HTML}{8FBCBB}          
\definecolor{NordBrightCyan}{HTML}{88C0D0}    
\definecolor{NordBlue}{HTML}{81A1C1}          
\definecolor{NordBrightBlue}{HTML}{5E81AC}    
\definecolor{NordRed}{HTML}{BF616A}           
\definecolor{NordOrange}{HTML}{D08770}        
\definecolor{NordYellow}{HTML}{EBCB8B}        
\definecolor{NordGreen}{HTML}{A3BE8C}         
\definecolor{NordMagenta}{HTML}{B48EAD}       

\begin{document}

\title{Generalized Krylov Complexity}  


\author{Amin Faraji Astaneh$^{1,2}$}
\author{Niloofar Vardian$^2$}
\email{faraji@sharif.ir}
\email{niloofar.vardian72@sharif.edu}
\affiliation{$^1$Department of Physics, Sharif University of Technology, P.O. Box 11155-9161, Tehran, Iran}
\affiliation{$^2$Research Center for High Energy Physics, Department of Physics, Sharif University of Technology, P.O. Box 11155-9161, Tehran, Iran}

\begin{abstract}
\noindent

We extend the concept of Krylov complexity to include general unitary evolutions involving multiple generators. This generalization enables us to formulate a framework for generalized Krylov complexity, which serves as a measure of the complexity of states associated with continuous symmetries within a model. Furthermore, we investigate scenarios where different directions of transformation lead to varying degrees of complexity, which can be compared to geometric approaches to understanding complexity, such as Nielsen complexity. In this context, we introduce a generalized orthogonalization algorithm and delineate its computational framework, which is structured as a network of orthogonal blocks rather than a simple linear chain. Additionally, we provide explicit evaluations of specific illustrative examples to demonstrate the practical application of this framework.
\end{abstract}
\maketitle

\section{Introduction}
The idea of quantum complexity has gained significant interest in quantum information science. Researchers from a variety of physics disciplines, ranging from quantum information theory to high-energy physics, have made concerted efforts to rigorously define this concept and calculate it for different quantum systems. The prevailing understanding is that quantum complexity quantifies the size of a quantum circuit or the load of a algorithm required to evolve an initial reference state into a desired quantum state through unitary operations.

One framework where this concept is rigorously defined is Krylov complexity \cite{parker2019universal}, commonly referred to as spread complexity  for quantum states rather than operators \cite{balasubramanian2022quantum}.

Based on this prescription, we examine a state under study as a result of the time evolution of an initial reference state $|\psi_t\rangle = e^{-iHt}|\psi_0\ra$. We gradually deviate from the initial state at time zero along a trajectory obtained by solving the Schrodinger equation. This gradual evolution can be expressed by expanding the time evolution operator in time. 
\be
|\psi_t\ra=\sum_{n=0}\frac{(-it)^n}{n!}|\psi_n\ra\ , \ |\psi_n\ra=H^n|\psi_0\ra\, .
\ee
Intuitively, the rays \( |\psi_n\rangle \) can form a vector space whose size is related to the distance between the final state \( |\psi_t\rangle \) and the initial reference state \( |\psi_0\rangle \). This brings the concept of complexity to mind.

To be more precise, constructing the building blocks within this framework necessitates the selection of only linearly independent vectors. This can be accomplished using the standard Gram-Schmidt orthogonalization procedure, which is referred to as the Lanczos algorithm in this context. By applying this method, we arrive at the Krylov basis, which constitutes a vector space known as the Krylov space. More precicely, one must consider 
\(
\text{Span}\{H^n|\psi_0\ra\}
\)
and apply the Gram-Schmidt procedure to obtain the corresponding orthonormal basis denoted by $|K_n\ra$. By collecting these vectors, one constructs the Krylov space of dimension \( d_{\CK} \leq d_\CH \), where \( \CH \) represents the entire Hilbert space. The desired state vector can now be expanded in terms of these basis as follows
\be
|\psi_t\ra=\sum_{n=0}^{d_\CK-1}\phi_n(t)|K_n\ra\, .
\ee
The complexity can now be quantified based on the distribution of basis vectors needed to construct the desired state. The more complex this distribution is, the more complex the state becomes.
This motivates us to define the complexity of state as follows
\be
\CC_{|\psi_t\ra}=\sum n|\phi_n(t)|^2\, .
\ee
The intuition behind this definition is that since \( |\phi_n(t)|^2 \) represents the probabilities of finding the evolved state in the \( n \)th Krylov basis element, the Krylov complexity can be understood as the expectation value of \( n \) weighted by these probabilities. This is similar to calculating an average depth in the Krylov chain, which measures how much the state has spread.

Krylov complexity has been widely explored in quantum chaotic systems \cite{FarajiAstaneh:2025rlc,balasubramanian2022quantum, balasubramanian2025quantum, tang2023operator, bhattacharjee2025krylov, rabinovici2022krylov, scialchi2024integrability, gill2024complexity, bhattacharya2024krylov, camargo2024spread, scialchi2025exploring, rabinovici2021operator, bhattacharjee2023krylov, hornedal2022ultimate, erdmenger2023universal, chapman2024krylov, baggioli2024krylov}, many-body systems, and conformal field theories \cite{bhattacharjee2022probing, Nandy:2023brt, Banerjee:2022ime}. Beyond chaos, it has proven instrumental in studying topological and quantum phase transitions \cite{afrasiar2023time,  pal2023time, Chakrabarti:2025hsb}, high-energy QFT \cite{Imani:2025etp, Alishahiha:2022anw, Vardian:2024fsp, Iizuka:2023pov, Iizuka:2023fba, Malvimat:2024vhr, Vasli:2023syq, Camargo:2022rnt, Kundu:2023hbk, Adhikari:2022whf, Avdoshkin:2022xuw, Dymarsky:2021bjq}, thermalization \cite{Alishahiha:2024rwm}, and open quantum systems \cite{bhattacharya2022operator, bhattacharjee2023operator, mohan2023krylov, bhattacharya2023krylov, bhattacharjee2024operator, carolan2024operator}.
 For a review that compiles many of these studies,
see \cite{nandy2025quantum, Rabinovici:2025otw, Baiguera:2025dkc}.

What we aim to investigate in this paper arises from a fair question: why could we not generalize this concept to other unitary evolutions rather than just time?

In this paper, we aim to provide a comprehensive answer to the question. We explore a fully general unitary transformation characterized by multiple generators and transformation parameters. Our goal is to extend the concepts of Krylov complexity and posibbly the Lanczos algorithm to accommodate these more general cases. The result of our work is the introduction of a Krylov network, instead of a Krylov chain. In doing so, we define generalized Krylov complexity and compute it for specific examples, focusing on both Abelian and non-Abelian Lie algebras.


\section{Generalized Krylov complexity}\label{GKCdefinition}
To develop our idea of generalizing Krylov complexity, we consider the most general case possible. In this context, we take an orthogonal set of $N$ Hermitian generators $\{T_i\}$ that form a general Lie algebra $[T_i,T_j]=if^k_{ij}T_k$. These operators can be designated as generators of unitary evolution operators, which evolve a given base state $|\psi_0\ra$ to
\be\label{state}
|\psi\{\alpha_k\}\ra = \exp\left(\sum_{k=1}^Ni\alpha_k T_k\right)|\psi_0\ra\, .
\ee 
In a common sense, this set of unitary operators architects a circuit through which one can build a ray $|\psi\{\alpha_k\}\rangle$ starting from an initial state. The main question is what the Krylov complexity of the final state is, which is naturally closely linked to the complexity of the resulting circuit.

In the following, we will quickly introduce the idea of how to compute the generalized Krylov complexity. However, we postpone the details and intermediate steps to the next section, where we will proceed algorithmically, starting from the reference state, step by step and block by block.

By expanding the right-hand side of \eqref{state}, a Krylov basis can be derived as a collection of blocks resulting from block decomposition as follows
\be
\CK=\bigoplus_{n=0}^{N_B-1}\CK_n\, ,
\ee
where each of $N_B$ Krylov blocks, labeled by $ \mathcal{K}_n$ is governed by the Gram-Schmidt orthonormalization of its associated span of rays which reads 
\be
\text{Span}\, \{T_1^{k_1}\cdots T_N^{k_N}|\psi_0\ra\big\vert k_1+\cdots +k_N=n\}\, .
\ee
such that $\mathcal{K}_n$ be orthogonal to all $\mathcal{K}_m, m< n$, see appendix \ref{AppA}.
In an appropriate representation
we denote each block as
\be
\CK_n=\text{Span}\{|K_n^i\ra,i=0,\cdots,d_{\CK_n}-1\}\,,
\ee
where $|K_n^i\ra$ belongs to the orthonormalized Krylov basis of block $n$, which forms a vector space of dimension $d_{\CK_n}$.
Now the generic state $|\psi\{\alpha_k\}\ra$ can be expanded in these basis as
\be
|\psi\{\alpha_k\}\ra=\sum_{n=0}^{N_B-1}\sum_{i=0}^{d_{\CK_n}-1} \phi_n^i(\{\alpha_k\})|K_n^i\ra\, .
\ee
The dimension of the Krylov space as a whole equals to $d_\CK=N_B\, d_{\CK_n}$.

We propose then the most natural generalization of the standard Krylov complexity as follows
\be\label{GCK}
\CC_{|\psi\{\alpha_k\}\ra}=\sum_{n=0}^{N_B-1}\sum_{i=0}^{d_{\CK_n}-1}n\vert \phi_n^i(\{\alpha_k\})\vert^2\, .
\ee
Note that for the standard Krylov complexity, there are $N_B$ blocks of dimension $d_{\CK_n}=1$. Furthermore, since all vectors within a given block are linear combinations of vectors generated by applying the same number of generators to the initial state, we can assume that they are equidistant from the initial state. Thus, only the index $n$ should be present as a coefficient in the definition of generalized Krylov complexity, and the order in which the Gram-Schmidt procedure is applied within each block is irrelevant.

It is important to note that when we generalize the Krylov complexity, we do not necessarily mean that we are only enlarging the basis space. In fact, the resulting circuit might not be homogeneous in its generator space. Instead, one might define a weighted generalized Krylov complexity by assigning a weight function to the Krylov space. In this sense, some blocks may contribute more effectively in preparing a state rather than others. To incorporate this, we start with
\be
\text{Span}\, \{T_1^{k_1}\cdots T_N^{k_N}|\psi_0\ra\big\vert \mu_1k_1+\cdots +\mu_nk_N=n\}\, ,
\ee
to define the weighted Krylov blocks. The distribution of \( \mu_k \) values defines the geometry of a circuit as a union of blocks, which requires further investigation. An important question then is which distribution of the weight function leads to an optimal state preparation. We will address this in the following sections.

A crucial step in calculating the generalized Krylov complexity is designing an algorithm to identify the Krylov blocks and to cover the Krylov space. This process is what we refer to as the generalized orthogonalization algorithm. The following section is dedicated to this.

\section{Generalized Orthogonalization Algorithm} \label{GOrtho}
In this section, we aim to generalize the orthogonalization algorithm for the problem of general unitary evolution. However, due to the complexity involved, we will not present the algorithm for deriving the Lanczos coefficients in its most general form. Instead, we will focus on writing an algorithm based on the projection onto the different blocks.

We begin with the initial state $\ket{\psi_0}$ and have a set of generators $\{ T_k, k=1,...,N \}$. For now, we assume that all generators are equally weighted.

What we require is to determine the projections onto different Krylov blocks, step by step.

The $0$-th block is generated from the initial state and is always one-dimensional, represented as 
\begin{equation}
\mathcal{K}_0 = \text{span} \{ \ket{K_0^0} = \ket{\psi_0} \}\, .
\end{equation}
The projection onto \( \mathcal{K}_0 \) is denoted by 
\begin{equation}
P_0 = \ket{K_0^0} \bra{K_0^0} = \ket{\psi_0} \bra{\psi_0}\, .
\end{equation}
In the next step, we need to construct the first block from the set of states given by 
\begin{equation}
\{ T_1 \ket{\psi_0}, T_2 \ket{\psi_0}, \cdots, T_N \ket{\psi_0} \}\, ,
\end{equation}
such that \( \mathcal{K}_1 \perp \mathcal{K}_0 \). To proceed with building \( \mathcal{K}_1 \), we first need to remove the images of all vectors in the above set from the subspace \( \mathcal{K}_0 \). For each vector \( \ket{v} \), this can be achieved by replacing it with the vector 
\begin{equation}
\ket{v'} = (I - P_0) \ket{v}\, ,
\end{equation}
since \( P_0 \ket{v'} = 0 \). Consequently, the first block can be constructed from the set 
\begin{equation}
\text{Span} \{ (I - P_0) T_1 \ket{\psi_0}, \cdots, (I - P_0) T_N \ket{\psi_0} \}.
\end{equation}
However, these vectors are not yet orthonormal. Therefore, we need to apply the Gram-Schmidt orthogonalization procedure to obtain 
\begin{equation}
\mathcal{K}_1 = \text{Span} \Big\{ \ket{K_1^0}, \ket{K_1^1}, \cdots, \ket{K_1^{d_{\mathcal{K}_1}-1}} \Big\}\, .
\end{equation}
For a generic \( \mathcal{K}_n \), we will similarly apply the Gram-Schmidt orthogonalization to 
\begin{equation}\label{T1TN}
\text{Span}\{(I - \sum_{i=0}^{n-1} P_i) T_1^{k_1} T_2^{k_2} \cdots T_N^{k_N} \ket{\psi_0} \}\, ,
\end{equation}
where \( k_1 + k_2 + \cdots + k_N = n \) and \( P_i \) is the projection onto the block \( \mathcal{K}_i \). This process yields the generalized Krylov basis \( \ket{K_n^i} \), ensuring that the blocks remain orthonormal to each other.

We must continue this procedure until we reach the point where 
\begin{equation}
    \mathcal{K}_i = 0, ~~~~ \forall i \geq N_B\, .
\end{equation}
Finally, the generalized Krylov complexity can be derived from equation \eqref{GCK}, where 
\(\phi_n^i(\{\alpha_k\}) = \langle K_n^i | \psi\{\alpha_k\} \rangle\).
We can express the complexity defined in \eqref{GCK} as 
\begin{equation}
    \begin{split}
      \CC_{|\psi\{\alpha_k\}\rangle} =& \sum_{n=0}^{N_B-1}\sum_{i=0}^{d_{\CK_n}-1} n \langle \psi\{\alpha_k\} | K_n^i \rangle \langle K_n^i | \psi\{\alpha_k\} \rangle \\
      =& \langle \psi\{\alpha_k\} | \sum_{n=0}^{N_B-1} n P_n | \psi\{\alpha_k\} \rangle\, ,
    \end{split}
\end{equation}
where \( P_n = \sum_{i=0}^{d_{\CK_n}-1} | K_n^i \rangle \langle K_n^i |\) is the projection onto the $n$-th block. Therefore, the generalized Krylov complexity can be interpreted as the expectation value of the generalized Krylov operator defined as 
\begin{equation}
    \hat{K}_{|\psi_0\rangle} = \sum_{n=0}^{N_B-1} n P_n\, ,
\end{equation}
on the state \( | \psi\{\alpha_k\} \rangle \).

It is worth noting that, without using the Gram-Schmidt procedure, one can alternatively determine the projection onto the subspace \( V = \text{span} \{ \ket{v_i} \} \) as follows, one can see \cite{Bahiru:2022ukn} for more detail.
First, we introduce the metric tensor defined by 
\be\label{metric}
G= [g_{ij}] = [\langle v_i | v_j \rangle]\, ,
\ee
which has an inverse denoted as \( G^{-1}=[g^{ij}] \). The projection onto the subspace \( V \) can then be expressed as
\begin{equation}\label{Pmetric}
    P_V = \sum_{i,j} g^{ij} \ket{v_i}\bra{v_j} \, .
\end{equation}
In our case, to find the projection onto \( \CK_n \), we can define the metric tensor for each block as 
\begin{multline}
   G_n= [ g(n)_{\{k_i\};\{k_i'\}}] = \\ [\bra{\psi_0} T_N^{k_N} \ldots T_1^{k_1} \left(I - \sum_{i=0}^{n-1} P_i\right) T_1^{k_1'} \ldots T_N^{k_N'} \ket{\psi_0} ]\, ,
\end{multline}
where \( n = \sum k_i = \sum k_i' \). Therefore, the projections in our problem can be derived from equation \eqref{Pmetric} by substituting equation \eqref{T1TN} for the constituent vectors. And finally, the generalized Krylov operator reads as 
\begin{multline}
    \hat{K}_{\ket{\psi_0}} = \sum_{n=0} ^{N_B-1} n  \sum_{\{k_i\};\{k_i'\}}  \\ g(n)^{\{k_i\};\{k_i'\}}~ 
    T_1^{K_1} ... T_N^{k_N} \ket{\psi_0}\bra{\psi_0} T_N^{k_N'}...T_1^{k_1'}.
\end{multline}
where $ G^{-1}_n = [g(n)^{\{k_i\};\{k_i'\}}]$. It is important to note that if there is redundancy in the set described by equation \eqref{T1TN}, we must restrict our analysis to the subspace orthogonal to the kernel of \( G_n \). Lastly, one can find the generalized Krylov operator directly using the form of the projectors obtained here. Additionally, we point out that the procedure of finding the blocks is repeated in this approach until reaching the point where \( G_i = 0 \) for all \( i \geq N_B \). And, the dimension of the entire Krylov basis can be found as 
\begin{equation}
    d_\CK = \sum_{i=0}^{N_B-1} d_{\CK_i} = \sum_{i=0}^{N_B-1} \tr P_i.
\end{equation}



\section{Generalized Krylov complexity AND SYMMETRIES OF THE SYSTEMS; concrete examples}

The generalized Krylov complexity presented in this work offers a valuable framework for analyzing the evolution of complexity under certain symmetries of a model. When a model is invariant under continuous transformations, the infinitesimal generators of these symmetries form a Lie algebra. Consequently, Lie algebras provide a natural basis for classifying and exploring symmetries across various fields of physics, including quantum mechanics, high-energy theory, and condensed matter systems.

In our framework, if \( T_k \) are the infinitesimal generators of a symmetry transformation in a model, our proposed generalized quantity measures the differences in complexity among the various states of the model under this symmetry transformation. This approach is applicable to both Abelian and non-Abelian Lie algebras.

To illustrate how our proposed mechanism operates in its most fundamental form, specifically within Abelian groups, we focus on the Abelian subgroups of $SO(n)$ characterized by the structure $U(1) \times U(1)$. Consequently, our aim is to determine the complexity of a state represented as
\be
U(\theta_1, \theta_2) \ket{\psi_0} = e^{i\theta_1 T_1 + i\theta_2 T_2} \ket{\psi_0}\,.
\ee
To accomplish this, we appropriately select the matrices as follows
\be
\begin{split}
    & T_1 = \text{diag}(a_1, \ldots, a_n)\,, \\
    & T_2 = \text{diag}(b_1, \ldots, b_n)\,.
\end{split}
\ee
Additionally, we can take $ \ket{\psi_0} = \frac{1}{\sqrt{n}}(1\  \ldots\ 1)\,$
which is a vector of dimension $n$ \footnote{In general, this construction applies to any initial state in an 
$n$-dimensional vector space. However, because the group is Abelian and its generators commute, we can choose a basis in which the generators are diagonal. In this basis, we can also express the matrix elements of the initial state. If the initial state has some zero entries in this basis, it is equivalent to projecting the full vector space onto the subspace where the state has non-zero components. This effectively reduces the problem to working within a smaller Hilbert space from the outset. }. We then proceed to compute the generalized Krylov complexity, analyzing the Krylov blocks step by step. One can also extend the discussion above to the  bigger group of $U(1) ^d$ with $d$ commutative generators. 

The simplest example involves two independent rotations in two $\mathbb{R}^2$ planes. We consider this symmetry transformation as a subgroup of the rotation group $SO(4)$, parameterized by two angles, $\theta_1$ and $\theta_2$, acting in two orthogonal planes. The detailed calculations are provided in the appendix \ref{appabelian}. Here, we only report the final simple result for the generalized Krylov complexity, which reads
\be
\mathcal{C}_\psi(\theta_1, \theta_2) = 1 - \cos\theta_1 \cos\theta_2\, .
\ee
Beside this simple example, to ensure our case is sufficiently illustrative and involves the computation of a significant number of blocks, it is best to focus on a large \( n \) case. For this reason, we consider the Abelian subgroup  $U(1)_Y \times U(1)_{B-L}$ of the $SO(10)$ group, which indeed holds a real place in Grand Unified Theories (GUTs), see Fig. \ref{subfig21}. In appendix \ref{appabelian}, one can find the detail of the result for this example. 

As a non-Abelian symmetry group we can consider the $SU(2)$ group. 
In the $J$ representation of $SU(2)$, any initial state $\ket{\psi_0}$ transforms under a general $SU(2)$ transformation as 
\begin{equation}
    \ket{\psi (\theta_x, \theta_y, \theta_z)} = \exp \big( -i \vec{\theta}\cdot\vec{J}\big) \ket{\psi_0}\, .
\end{equation}
The complexity of \( \ket{\psi (\theta_x, \theta_y, \theta_z)} \) in comparison with \( \ket{\psi_0} \) can be computed using the generalized Krylov complexity defined here for the generators \( J_x, J_y, \) and \( J_z \). More details about this class of symmetry group can be found in appendix \ref{abbnonabelian}. In particular, if \( \ket{\psi_0} = \ket{j,m} \), the generalized Krylov complexity can be obtained from \eqref{jm}. For the case of \( \ket{j,j} \), it reduces to
\begin{equation}
     C_\psi(\theta_x, \theta_y, \theta_z) = 2j~ \sin^2 \Big(\frac{|\theta|}{2} \Big) \frac{|\theta|^2 - \theta_z^2}{|\theta|^2}\, .
\end{equation}

\begin{figure}[h!]
    \centering
    \begin{subfigure}[b]{0.35\textwidth}
        \centering
        \includegraphics[width=\linewidth]{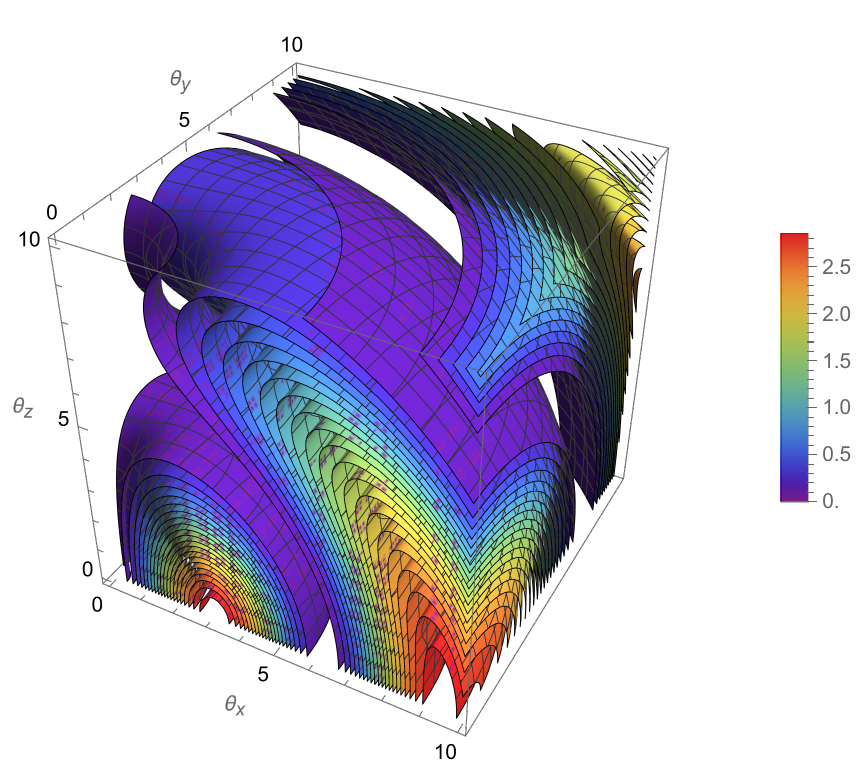}
        \caption{initial state is  $ \ket{j=3/2,m=3/2}$}
        \label{subfig11}
    \end{subfigure}

    \begin{subfigure}[b]{0.35\textwidth}
        \centering
        \includegraphics[width=\linewidth]{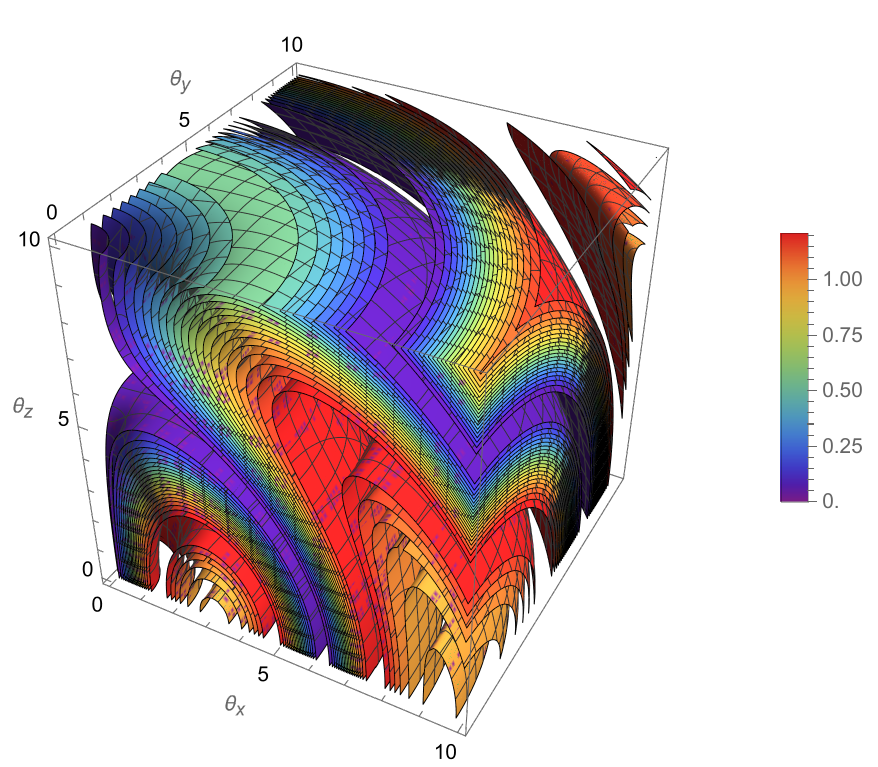}
        \caption{initial state is  $ \ket{j=3/2,m=1/2}$}
        \label{subfig12}
    \end{subfigure}

    \caption{The complexity of the initial state in the $ j= 3/2$ representation of $ SU(2)$, under the evolution by the unitary evolution of the algebra $SU(2)$ as a function of $ \theta_x, \theta_y, \theta_z$.}
    \label{fig1}
\end{figure}

In Fig. \eqref{fig1}, one can observe the results for the generalized Krylov complexity of the initial states \( \ket{j=3/2,m=3/2} \) and \( \ket{j=3/2,m=1/2} \) under the evolution by \( SU(2) \) algebra.
\\

\section{Weighted Generalized Krylov complexity}

The generalized Krylov complexity defined in Sec. \ref{GKCdefinition} fails to recognize that some unitary transformations are more difficult to implement than others, as it does not discriminate between different directions in the manifold of unitaries. In other words, moving in the direction of \( T_i \) might be more complex than moving in the direction of \( T_j \).

This idea is explicitly integrated into the Nielsen complexity framework \cite{nielsen2006quantum, brown2019complexity, nielsen2005geometric, dowling2006geometry}. In this geometric approach to quantum complexity, the central concept is that moving in certain directions on the manifold of unitary operators is more complex than moving in others. A key feature of this framework is the definition of a cost functional along a path in the space of unitaries or states, which assigns different weights to various directions, making some paths more costly than others. In this section, we aim to address this question in the context of generalized Krylov complexity.

To distinguish between different generators, we assign a factor of \( \mu_i \) to each generator \( T_i \). For a state \( U(\{\alpha_i\}) \ket{\psi_0} \) on the manifold of the corresponding Lie algebra, we need to block-decompose the Krylov basis and assign a number to each block, in contrast to \( n \) (where \( n = k_1 + k_2 + \ldots + k_N \)) in the non-weighted case. It is important to note that these parameters should be increasing; as we apply a fixed generator more frequently, the complexity of the state should increase.
In terms of the factors \( \mu_i \), we can define these parameters as follows
\begin{equation}
   \tilde{n} := \sum _{i=1} ^N  \mu_i k_i\, ,
\end{equation}
and one can choose a normalization for the factors $\mu_i$ such that $\sum _{i=1} ^N  \mu_i =N $.

Now, to find the Krylov blocks and calculate the generalized Krylov complexity, one must first sort all possible values for $\tilde{n}$ while ensuring that $k_i \in \mathbb{N}$. Subsequently, these values are considered in ascending order until the entire Krylov space is covered.

For a fixed $\tilde{n}$, we need to construct the corresponding Krylov block $\CK_{\tilde{n}}$ by orthogonalizing the set of vectors given by
\begin{equation}
    \{ T_1^{k_1} \ldots T_N^{k_N}~ \ket{\psi_0} | \sum_{i=1}^N  \mu_i k_i = \tilde{n} \}\, .
\end{equation}
It is required that $ \CK_{\tilde{n}} \perp  \CK_{\tilde{m}}$ for all $\tilde{m} < \tilde{n}$. By applying the Gram-Schmidt procedure, one can determine the block decomposition of the Krylov space as $ \CK = \bigoplus_{\tilde{n}} \CK_{\tilde{n}}$, where
\begin{equation}
    \CK_{\tilde{n}} = \text{Span} \{ \ket{K_{\tilde{n}}^i}, i=0,\ldots, d_{\CK_{\tilde{n}}}-1 \}\, .
\end{equation}
The weighted generalized Krylov complexity can then be expressed as 
\begin{equation}
    C_{\ket{\psi\{\alpha_k\}}} = \sum_{\tilde{n}} \sum_{i=0}^{d_{\CK_{\tilde{n}}}-1}~ \tilde{n}~ | \phi_{\tilde{n}}^i ( \{\alpha_k\})|^2\, ,
\end{equation}
where $ \phi_{\tilde{n}}^i (\{\alpha_k\}) = \langle K_{\tilde{n}} ^i | \psi \{ \alpha_k\}\rangle$. The generalized orthogonalization algorithm discussed in Sec. \ref{GOrtho} can be naturally extended to the weighted cases.

\begin{figure}[h!]
    \centering
    \begin{subfigure}[b]{0.35\textwidth}
        \centering
        \includegraphics[width=\linewidth]{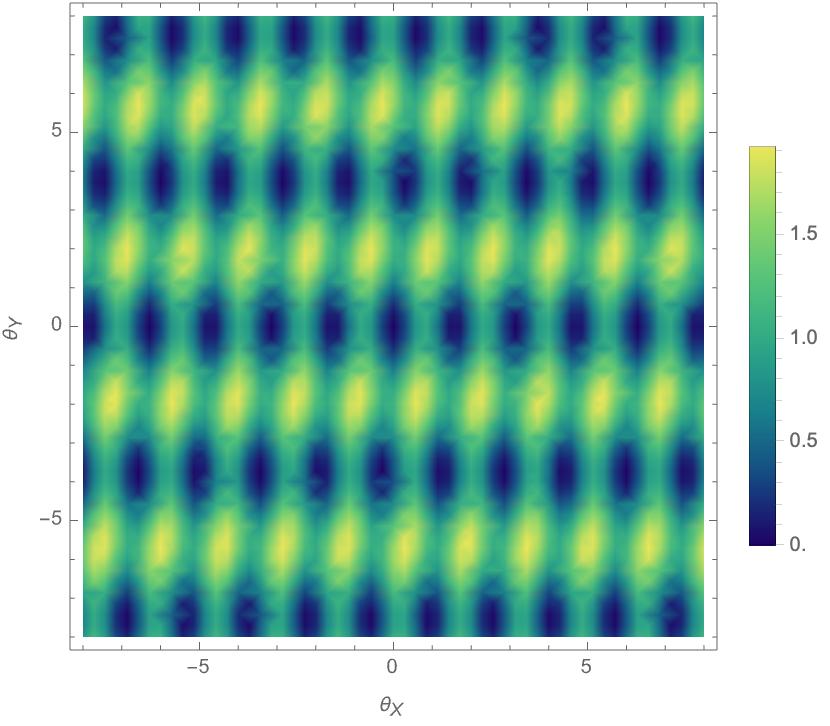}
        \caption{$\mu_X = \mu _Y =1$ }
        \label{subfig21}
    \end{subfigure}

    \begin{subfigure}[b]{0.35\textwidth}
        \centering
        \includegraphics[width=\linewidth]{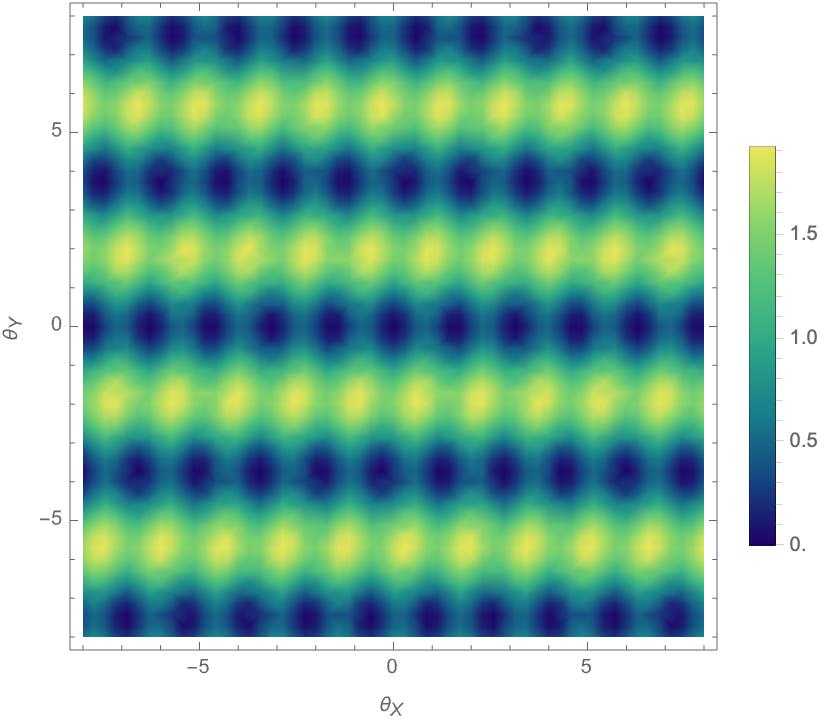}
        \caption{$\mu_X = 1/2,  \mu _Y =3/2$ }
        \label{subfig22}
    \end{subfigure}

    \caption{The Krylov complexity of the initial state \eqref{iniGUT} under the evolution by the $U(1)_X \times U(1)_Y$}
    \label{fig2}
\end{figure}

As a simple example, we revisit the scenario of two independent rotations in four dimensions, this time with two unequal weights. For this case, we select $\mu_1 = \frac{1}{2}$ and $\mu_2 = \frac{3}{2}$. The details for calculating the weighted complexity are provided in appendix \ref{appabelian}. Here, we quote the final result, which is given by
\begin{equation}
\tilde{\mathcal{C}}_{\psi} (\theta_1, \theta_2) = \frac{1}{4} \left( 3 - 2 \cos \theta_1 \cos \theta_2 - \cos 2 \theta_2 \right)\, .
\end{equation}
It is evident that the previous symmetry between the two rotations has been disrupted due to the unequal weights assigned in the corresponding directions. In reference to the more complex example introduced earlier, as shown in Fig. \ref{fig2}, we can compare the generalized Krylov complexity results for the \(U(1)_Y \times U(1)_{B-L}\) group in both weighted and non-weighted cases. A more detailed discussion of these cases can be found in appendix \ref{appabelian}.

\section{Discussion}
In this paper, we have expanded the concept of Krylov complexity to include a general unitary transformation with multiple generators. Our goal has been to demonstrate, inspired by Krylov complexity, how complex a state can become when constructed from a completely general unitary transformation of a reference state. Our analysis is conducted using a fully general Lie algebra, allowing us to measure the difference in complexity between states related by the continuous symmetry of the corresponding Lie algebra in a model. We have introduced the generalized Krylov complexity in a broad context and tested it with specific Abelian and non-Abelian groups. Additionally, we explore an interesting generalization by considering weighted complexity, taking into account different weights and contributions in various directions of transformation.

Specifically, by utilizing this weighted complexity, one can develop a geometric representation of Krylov complexity. A general unitary transformation in the parameter space, which acts in a weighted manner along various directions, resembles the concept of a quantum circuit that enables the preparation of a general state from a reference state. While this connection appears intuitive and is closely related to established concepts of circuit complexity, enhancing its precision could yield particularly interesting insights. See \cite{nielsen2006quantum, brown2019complexity, nielsen2005geometric, dowling2006geometry, chapman2018toward, hackl2018circuit, chapman2019complexity, caputa2019quantum, guo2018circuit, caputa2022geometry, lv2024building} for relevant references.

Moreover, our approach provides a natural generalization of Krylov complexity to open quantum systems described by Lindblad dynamics \cite{liu2023krylov, mohan2023krylov, bhattacharya2024speed}. When the Lindblad master equation can be formulated using the generators of a Lie algebra, this structure offers a cohesive perspective on operator growth and complexity within dissipative environments.
Our approach may facilitate the exploration of previously unexamined Krylov complexity in systems with time-dependent Hamiltonians, which is an important survey. The concept can be naturally applied to cases where the time-dependent Hamiltonian can be expressed as a linear combination of the generators of a Lie algebra. It is essential to consider that all coefficients $\alpha_i$ are functions of time, which means that the generalized Krylov complexity can be expressed as a function of time as well. It is important to be cautious about the time ordering in the definition of time-evolved states when calculating complexity. Furthermore, our approach can be extended to cases where $H = f(\{T_i\})$, allowing us to analyze a broader class of models. For a similar discussion on using Lie algebras to address the complexity of time-dependent Hamiltonians,one can see \cite{chowdhury2024krylov}. We plan to explore these topics in our future research.

We also note the potential to interpret the generalized Liouvillian operator from two Hamiltonians, such as in a two-sided AdS black hole with matter, where it is essential to clarify how to implement evolution with respect to the asymptotic boundaries. This interpretation has been explored in some papers, including \cite{aguilar2025geometry, aguilar2025chords, ambrosini2025operator, aguilar2025chords}, using this framework.


\begin{acknowledgements}
The authors would like to thank A. Azatov, B. Bajc, H. Georgi and S. Raby for usefule discussions.
\end{acknowledgements}

\bibliography{refs.bib} 

\onecolumngrid
\clearpage
\appendix


\section{Krylov Blocks}\label{AppA}

In Section \ref{GKCdefinition}, we consider an orthonormal set of \( N \) Hermitian generators \( T_i \) such that the trace satisfies \( \Tr[T_i^\dagger T_j] = \delta_{ij} \). These generators form a general Lie group, characterized by the commutation relation \( [T_i, T_j] = i f_{ij}^k T_k \). We begin with a generic initial state \( \ket{\psi_0} \) located at the origin of the manifold of vectors, which are derived from the initial state through the unitary evolution associated with the corresponding Lie group
\begin{equation}
    \ket{\psi\{\alpha_k\}} = \exp\left(\sum_{k=1}^Ni\alpha_k T_k\right) \ket{\psi_0}\, .
\end{equation}
We can expand the evolved state by using the Taylor expansion as 
\begin{equation}
    \begin{split}
     \ket{\psi\{\alpha_k\}} = & \sum _{n=0}^{\infty} \frac{i^n}{n!} \Big( \sum_k \alpha_k T_k\Big)^n \ket{\psi_0}
        \\
       = & \sum _{n=0}^{\infty}~~ \sum _{k_{i_1} +...+ k_{i_N} =n} \frac{i^n}{n!}~ \alpha _{i_1}^{k_{i_1}}... \alpha _{i_N}^{k_{i_N}}~ T _{i_1}^{k_{i_1}}...T _{i_N}^{k_{i_N}}~ \ket{\psi_0}\, .
    \end{split}
\end{equation}
Using the commutation relation between the generators of the algebra, one can rewrite it as 
\begin{equation}
     \ket{\psi\{\alpha_k\}} = \sum _{n=0}^{\infty}~~ \sum _{k_{1} +...+ k_{N} =n} \frac{i^n}{n!}~ A(k_1,...,k_N)~\alpha _{1}^{k_{1}}... \alpha _{N}^{k_{N}}~ T _{1}^{k_{1}}...T _{N}^{k_{N}}~ \ket{\psi_0}
\end{equation}
where $A(k_1,...,k_N) $ comes from the commutation relation between the generators.

As an illustration, consider a scenario involving three generators, denoted as $(T_1, T_2, T_3)$. At the $n=0$ term in the Taylor expansion, we have only the state $\ket{\psi_0}$, which contributes to $\CK_0$. Proceeding to the next term for $n=1$, we obtain the set of states $\{ T_1\ket{\psi_0}, T_2\ket{\psi_0}, T_3\ket{\psi_0}\}$. The first Krylov block can be constructed by applying the Gram-Schmidt procedure to this set, ensuring that the resulting states are orthogonal to $\CK_0$.

For the next term corresponding to $n=2$ in the Taylor expansion of the evolved state, we need to determine the set of states that will form the second Krylov block. To do this, we only need to consider the following terms
\be
\{ T_1^2 \ket{\psi_0}, T_1 T_2 \ket{\psi_0}, T_1 T_3 \ket{\psi_0}, T_2^2 \ket{\psi_0}, T_2 T_3 \ket{\psi_0}, T_3^2 \ket{\psi_0}\}\, .
\ee
It is essential that these terms are orthogonal not only to $\CK_0$ but also to $\CK_1$. For example, the vector $T_2 T_1 \ket{\psi_0}$ need not be evaluated separately, as it can be expressed as
\be
T_2 T_1 \ket{\psi_0} = T_1 T_2 \ket{\psi_0} - i f_{12}^3 T_3 \ket{\psi_0}\, .
\ee
Here, $T_1 T_2 \ket{\psi_0}$ is already included in our set of vectors, while $T_3 \ket{\psi_0}$ belongs to $\CK_1$. Hence, we can confidently remove it during the orthogonalization process.
Consequently, to construct the $n$-th block $\CK_n$, it suffices to consider the set of vectors given by
\be
\text{Span}\, \{T_1^{k_1} \cdots T_N^{k_N} \ket{\psi_0} \mid k_1 + k_2 + \cdots + k_N = n \}\, .
\ee
We derive $\CK_n$ through the Gram-Schmidt procedure applied to this set of vectors while maintaining the criterion that $\CK_n \perp \CK_m$ for all $m < n$. For more detailed information regarding the construction of the various Krylov blocks, refer to section \ref{GOrtho}.

\section{Examples}\label{appexample}

\subsection{Abelian Algebra, $ U(1)\times U(1)$}\label{appabelian}

In this section, we analyze the Krylov complexity under the action of an abelian group, the elements of $U(1) \times U(1)$. 

In the case of two rotations in two $\mathbb{R}^2$ planes, the transformation matrices are defined as follows
\be
T_1 =
\begin{pmatrix}
0 & -i & 0 & 0 \\
i & 0 & 0 & 0 \\
0 & 0 & 0 & 0 \\
0 & 0 & 0 & 0 
\end{pmatrix}
\quad \text{and} \quad
T_2 =
\begin{pmatrix}
0 & 0 & 0 & 0 \\
0 & 0 & 0 & 0 \\
0 & 0 & 0 & -i \\
0 & 0 & i & 0 
\end{pmatrix}\, .
\ee
The initial state vector is chosen as
\be\label{ini44}
\ket{\psi_0} = \frac{1}{2}(1\ 1\ 1\ 1)\, .
\ee
By applying the machinery for computing the generalized Krylov complexity, we find that
\be
d_{\mathcal{K}_0} = 1, \quad d_{\mathcal{K}_1} = 2, \quad d_{\mathcal{K}_2} = 1, \quad d_{\mathcal{K}_{n \geq 3}} = 0\, .
\ee
After a straightforward calculation, the Krylov operator is determined to be
\be
\hat{K}_{\ket{\psi_0}} = \frac{1}{2}
\begin{pmatrix}
2 & 0 & -1 & -1 \\
0 & 2 & -1 & -1 \\
-1 & -1 & 2 & 0 \\
-1 & -1 & 0 & 2 
\end{pmatrix}\, .
\ee
Thus, the complexity function can be expressed as
\be
\mathcal{C}_\psi(\theta_1, \theta_2) = 1 - \cos\theta_1 \cos\theta_2\, .
\ee

\begin{figure}[h!]
    \centering
    \begin{subfigure}[b]{0.43\textwidth}
        \centering
        \includegraphics[width=\linewidth]{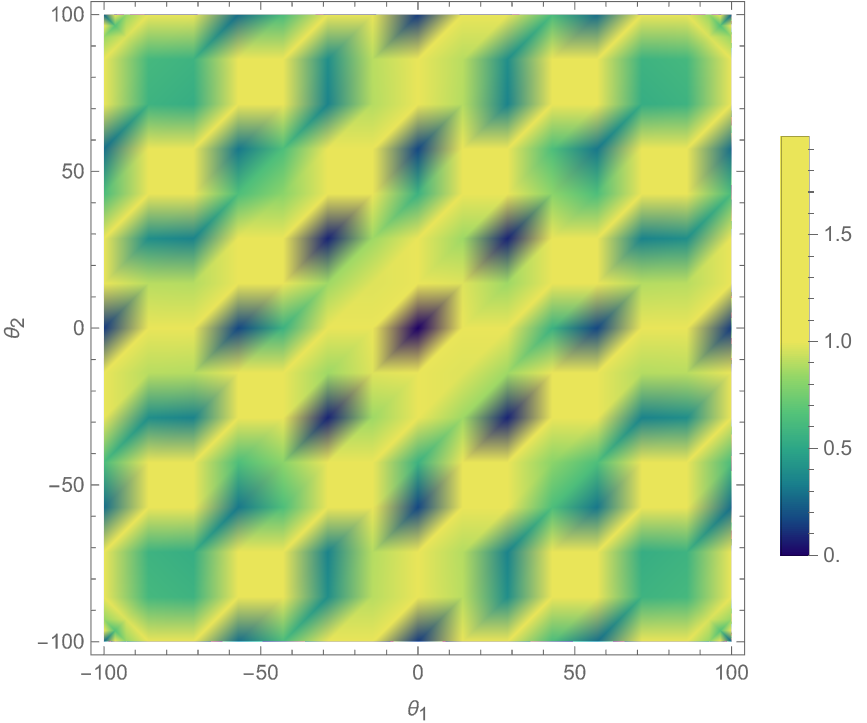}
        \caption{Two generators weighted equally.}
        \label{fig:subfig1000}
    \end{subfigure}
    \hfill

    \begin{subfigure}[b]{0.43\textwidth}
        \centering
        \includegraphics[width=\linewidth]{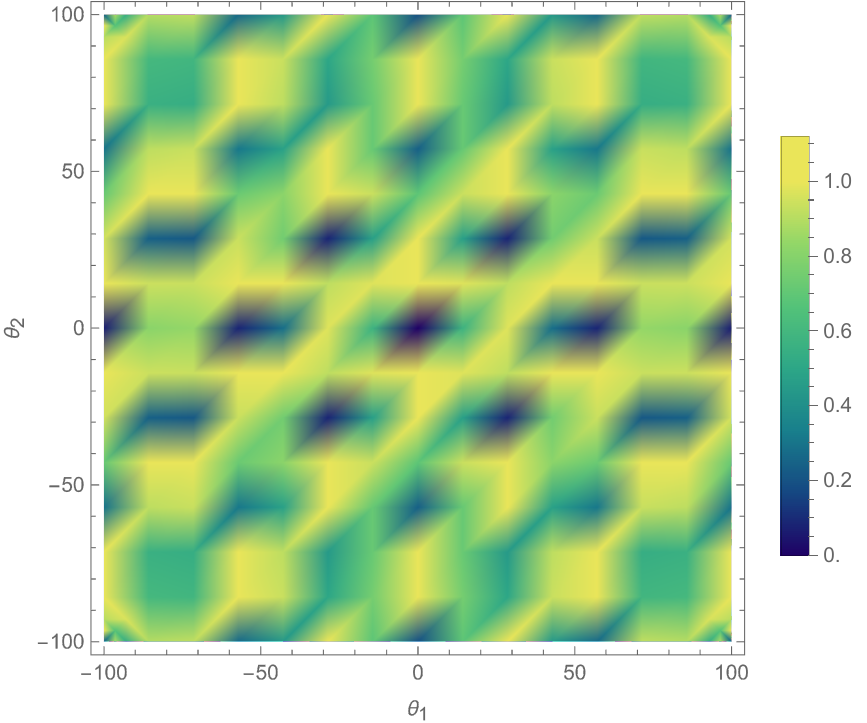}
        \caption{$\mu_1 =1/2, \mu_2 = 3/2 $}
        \label{fig:subfig2000}
    \end{subfigure}
    \begin{subfigure}[b]{0.43\textwidth}
        \centering
        \includegraphics[width=\linewidth]{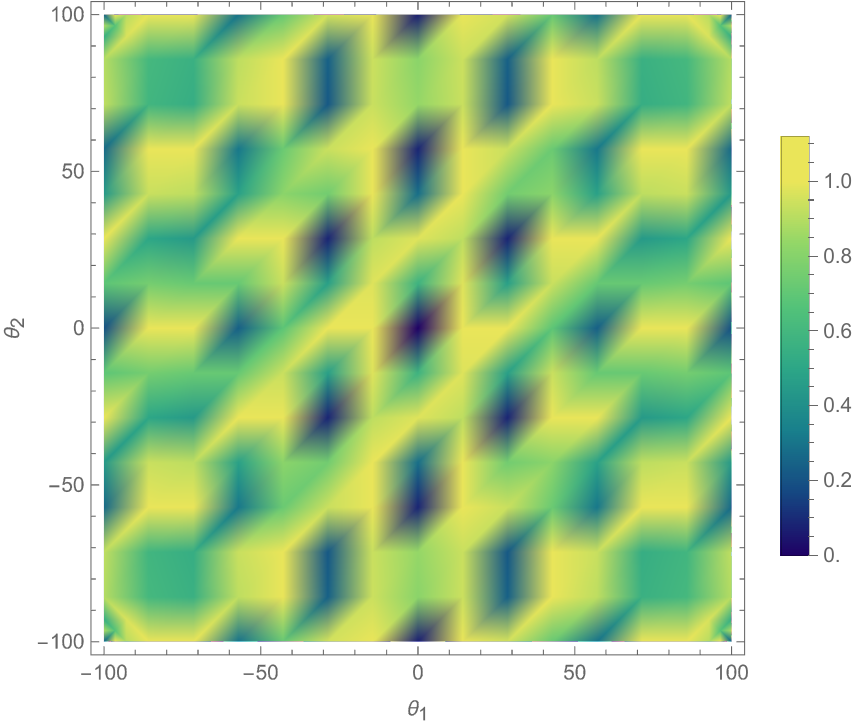}
        \caption{$\mu_1 =3/2, \mu_2 = 1/2 $}
        \label{fig:subfig3000}
    \end{subfigure}

\caption{The generalized Krylov complexity of the initial state \eqref{ini44} under the evolution by the two rotations as a $ U(1) \times U(1)$ subgroup of $SO(4)$ as a function of $ \theta_1$ and $ \theta_2$.}
    \label{Figure99}
\end{figure}

Now, let us consider that moving in the direction of these two generators does not create the same amount of complexity. First, let us consider the corresponding weights for the generators to be
\begin{equation}
    \mu_1 = \frac{1}{2}, ~~~~~~~  \mu_2 = \frac{3}{2}\, . 
\end{equation}
As a result, the Krylov blocks can be obtained as
\begin{equation}
    \begin{split}
       \tilde{\CK}_0 =& \text{span}\{ \ket{\psi_0} \}\, ,
        \\
         \tilde{\CK}_{1/2} =& \text{span}\{(I- \tilde{P}_0)T_1~ \ket{\psi_0} \}\, ,
        \\
        \tilde{\CK}_1 =& \text{span}\{(I-\tilde{P}_0-\tilde{P}_{1/2})T_1^2~ \ket{\psi_0} \}\, ,
        \\
         \tilde{\CK}_{3/2} =& \text{span}\{(I-\tilde{P}_0 -\tilde{P}_{1/2} -\tilde{P}_1)T_1^3~ \ket{\psi_0}, (I-\tilde{P}_0 -\tilde{P}_{1/2} -\tilde{P}_1)T_2~ \ket{\psi_0} \}\, ,
    \end{split}
\end{equation}
where
\be
d_{ \tilde{\CK}_0} = 1, \quad d_{ \tilde{\CK}_{1/2}} = 1, \quad d_{ \tilde{\CK}_1} = 1,\quad d_{ \tilde{\CK}_{3/2}} = 1\quad d_{ \tilde{\CK}_{n > 3/2}} = 0\, .
\ee
After a straightforward calculation, the weighted  Krylov operator is determined to be
\be
\hat{\tilde{K}}_{\ket{\psi_0}} =\frac{1}{4}
\begin{pmatrix}
2 & 0 & -1 & -1 \\
0 & 2 & -1 & -1 \\
-1 & -1 & 4 & -2 \\
-1 & -1 & -2 & 4 
\end{pmatrix}\, .
\ee
and the weighted generalized Krylov complexity can be expressed as 
\begin{equation}
   \tilde{ \mathcal{C}}_{\psi} (\theta_1, \theta_2) = \frac{1}{4} ( 3 -2 \cos \theta_1 \cos \theta_2 - \cos 2 \theta_2)\, .
\end{equation}
If we set the weights to \( \mu_1 = \frac{3}{2} \) and \( \mu_2 = \frac{1}{2} \), we need to swap \( \theta_1 \) and \( \theta_2 \) in the expression for the weighted generalized Krylov complexity. In Figure \ref{Figure99}, the plot of the generalized Krylov complexity as a function of \( \theta_1 \) and \( \theta_2 \) is displayed.
\\

As a more complicated example, we consider the $U(1)_X \times U(1)_Y$ subgroup of $SO(10)$ as a model for a grand unified theory (GUT) \cite{baez2010algebra, fritzsch1975unified,ref16}. It is crucial to highlight that the specific physical details of this model are not essential for our discussion. What is significant is the presence of a sufficiently large \( U(1) \times U(1) \) subgroup, which provides a relevant example to effectively illustrate the subtleties involved in our calculations.


In $\textbf{10}$ representation, one has \cite{Raby:1995uv,Bajc:2004xe}
\begin{equation}
    \begin{split}
        X &= 2\,\mathbb{I}_{5\times 5}\otimes \sigma_2\, ,
        \\
        Y&= \text{diag}\Big(\frac{2}{3},\frac{2}{3},\frac{2}{3},-1,-1\Big)\otimes \sigma_2\, ,
    \end{split}
\end{equation}
where $\mathbb{I}_{5\times 5}$ is the Identity matrix and $\sigma_2$ is the second Pauli matrix.
We also consider the initial state in this basis to be 
\begin{equation}\label{iniGUT}
    \ket{\psi_0}= \frac{1}{\sqrt{10}} (1\,1\,1\,1\,1\,1\,1\,1\,1\,1)\, .
\end{equation}
and we aim to study the generalized Krylov complexity of the states related to $\ket{\psi_0}$ by the unitary elements of the group $ U(1)_X \times U(1)_Y$, i.e.
\begin{equation}\label{GUTstate}
    \ket{\psi(\theta_X, \theta_Y)} = \exp [ i ( \theta_X X + \theta_Y Y)] \ket{\psi_0}\, .
\end{equation}
as a function of $\theta_X$ and $\theta_Y$.

First, we assume that both generators are weighted equally 
\begin{equation}
    \mu_X = \mu_Y=1\, .
\end{equation}
We start with 0-block and set $ \ket{K_{0,0}}= \ket{\psi_0}$, therefore
\begin{equation}
    \mathcal{K}_0 = \text{span} \{ \ket{K_{0,0}}= \ket{\psi_0}\}\, ,
\end{equation}
and
\begin{equation}
    P_0= \ket{\psi_0}\bra{\psi_0}\, .
\end{equation}
To construct the 1st block, we consider
\begin{equation}
    \mathcal{K}_1= \text{span}\{(I-P_0)X \ket{\psi_0},(I-P_0)Y \ket{\psi_0} \}\, ,
\end{equation}
where one can check that $ \tr (P_1) =2 $ which means
\begin{equation}
     \dim \mathcal{K}_1 =2\, .
\end{equation}
In other words, by doing the Gram-Schmidt in $\mathcal{K}_1$. one can reach two linearly independent vectors.
We need to continue and find other blocks as 
\begin{equation}
\begin{split}
     & \mathcal{K}_2= \text{span}\{(I-P_0-P_1)X^2 \ket{\psi_0},(I-P_0-P_1)XY \ket{\psi_0}, (I-P_0-P_1)Y^2 \ket{\psi_0} \} \, , \ \text{with}\ 
     \\
     & \dim \mathcal{K}_2 = \tr (P_2) =1\, .
\end{split}
\end{equation}

One can check that 
\begin{equation}
    \mathcal{K}_m =0\ \text{for} \ m \geq 3\, ,
\end{equation}
and the entire Krylov space is 
\begin{equation}
    \mathcal{K} = \oplus_{i=0}^{2} \mathcal{K}_i\, ,
\end{equation}
where
\begin{equation}
    \dim \mathcal{K} = \sum _{i=0}^2 \dim \mathcal{K}_i = \sum _{i=0}^2 \tr (P_i) = 1+2+1 =4 \, .
\end{equation}

Finally, the generalized Krylov complexity is 
\begin{equation}
    C_\psi (\theta_X, \theta_Y)= \bra{\psi(\theta_X, \theta_Y)} P_1 + 2 P_2 \ket{\psi(\theta_Y, \theta_{B-L})}\, ,
\end{equation}
and thus one can find
\begin{multline}\label{U1C}
      C_\psi (\theta_X, \theta_Y)=  \frac{1}{50} \left\{49 + 2 \cos\left(4\theta_X - 2\theta_Y\right) - 24 \cos\left(4\theta_X - \frac{\theta_Y}{3}\right) - 24 \cos\left(\frac{5\theta_Y}{3}\right) - 3 \cos\left[\frac{4}{3}(3\theta_X + \theta_Y)\right]\right\}\, .
\end{multline}
Figure \eqref{subfig21} shows the generalized Krylov complexity as a function of $(\theta_Y, \theta_{B-L})$. 

Now, let us assume that evolution along these $U(1)$ generators does not generate the same complexity. We may therefore weight them as follows
\begin{equation}
    \mu_X = \frac{1}{2}, \qquad\qquad \mu_Y =\frac{3}{2}\, .
\end{equation}
Here, the blocks labeled by 
\begin{equation}
    \tilde{n} = k_X \mu_X + k_Y \mu_Y = \frac{1}{2} k_X + \frac{3}{2} k_Y, \quad k_X, k_Y \in \mathbb{N}
\end{equation}
are sorted from 0 to higher values, corresponding to positive half-integer numbers.

As in the non-weighted case, the 0-block has $k_X = k_Y = 0$
\begin{equation}
   \tilde{\mathcal{K}}_0 = \text{span} \{ \ket{K_{0,0}} = \ket{\psi_0} \}\, ,
\end{equation}
with
\begin{equation}
    \tilde{P}_0 = \ket{\psi_0}\bra{\psi_0}\, , \ \text{with}\ \dim \tilde{\mathcal{K}}_0 = \tr(\tilde{P}_0) = 1\, .
\end{equation}
The first non-trivial block corresponds to the smallest $\tilde{n} > 0$, namely $\tilde{n} = 1/2$, obtained with $(k_X = 1, k_Y = 0)$
\begin{equation}
 \tilde{\mathcal{K}}_{1/2} = \text{span}\{(I-\tilde{P}_0)X \ket{\psi_0}\} \, , \ \text{with}\ \dim \tilde{\mathcal{K}}_{1/2} = \tr(\tilde{P}_{1/2}) = 1\, .
\end{equation}
The next block, with $\tilde{n} = 1$, is found by setting $(k_X = 2, k_Y = 0)$. At and beyond this level, the operator $X^k$ generates null states. Therefore, we truncate the expansion in $X$ here, omitting higher-order terms.

The subsequent value $\tilde{n} = 3/2$ can be achieved by considering $(k_X = 0, k_Y = 1)$, giving
\begin{equation}
\begin{split}
     & \tilde{\mathcal{K}}_{3/2} = \text{span}\{(I-\tilde{P}_0-\tilde{P}_{1/2})Y \ket{\psi_0}\}\, , \ \text{with}\ \\
     & \dim \tilde{\mathcal{K}}_{3/2} = \tr(\tilde{P}_{3/2}) = 1.
\end{split}
\end{equation}
For $\tilde{n} = 2$, we consider $(k_X = 1, k_Y = 1)$
\begin{equation}
\begin{split}
     & \tilde{\mathcal{K}}_2 = \text{span}\{(I-\tilde{P}_0-\tilde{P}_{1/2}-\tilde{P}_{3/2})XY \ket{\psi_0}\}\, , \ \text{with}\ \\
     & \dim \tilde{\mathcal{K}}_2 = \tr(\tilde{P}_2) = 1.
\end{split}
\end{equation}

For $\tilde{n} > 2$, we find $\tilde{P}_{\tilde{n}} = 0$, which means
\begin{equation}
   \tilde{\mathcal{K}}_{\tilde{n}} = 0 \quad \text{for} \quad \tilde{n} > 2\, .
\end{equation}
Thus, the Krylov space decomposes as
\begin{equation}
    \tilde{\mathcal{K}} = \bigoplus_{i \in I} \tilde{\mathcal{K}}_i, \quad I = \{0,1/2,3/2,2\}\, ,
\end{equation}
with total dimension
\begin{equation}
    \dim \tilde{\mathcal{K}} = \sum_{i \in I} \dim \tilde{\mathcal{K}}_i = \sum_{i \in I} \tr(\tilde{P}_i) = 4\, .
\end{equation}
The weighted generalized Krylov complexity of the state \eqref{GUTstate} is given by
\begin{multline}\label{U1CW}
    \tilde{C}_\psi(\theta_X,\theta_Y) = \bra{\psi(\theta_X, \theta_Y)} \sum_{i \in I} \tilde{n}_i \tilde{P}_{\tilde{n}_i} \ket{\psi(\theta_X, \theta_Y)} \\
    = \bra{\psi(\theta_X, \theta_Y)} \left( \frac{1}{2} \tilde{P}_{1/2} + \frac{3}{2} \tilde{P}_{3/2} + 2 \tilde{P}_2 \right) \ket{\psi(\theta_X, \theta_Y)}\\
    =\frac{1}{100} \left\{ 97 + 2 \cos\left(4\theta_X - 2\theta_Y\right) - 24 \cos\left(4\theta_X - \frac{\theta_Y}{3}\right) - 72 \cos\left(\frac{5\theta_Y}{3}\right) - 3 \cos\left[\frac{4}{3} \left(3\theta_X + \theta_Y\right)\right] \right\}\, .
\end{multline}

Figure \ref{subfig22} shows the weighted generalized Krylov complexity as a function of $(\theta_X, \theta_Y)$.

Comparing \eqref{U1CW} with \eqref{U1C} one finds that
\be
 \tilde{C}_\psi (\theta_X, \theta_Y)\big\vert_{(\mu_X=\frac{1}{2},\mu_Y=\frac{3}{2})}=\frac{1}{2} C_\psi (\theta_X, \theta_Y)\big\vert_{(\mu_X=\mu_Y=1)}+\frac{24}{25}\sin^2\left(\frac{5}{6}\theta_Y\right)\, .
\ee

\subsection{Non-Abelian Algebra, $SU(2)$}\label{abbnonabelian}

We now analyze an example where the symmetry generators form the $SU(2)$ algebra. We begin with the $SU(2)$ Lie algebra
\begin{equation}
    [J_i,J_j] = i \epsilon_{ijk}J_k\, ,
\end{equation}
where $J_1 = J_x$, $J_2 = J_y$, $J_3 = J_z$, satisfy the trace orthogonality condition $\Tr[J_i J_j] \propto \delta_{ij}$.
The ladder operators are defined as
\begin{equation}
    J_{\pm} = J_x \pm i J_y\, ,
\end{equation}
with 
\begin{equation}
    J_0 = J_z\, .
\end{equation}
This transforms the algebra to
\begin{equation}
    [J_0, J_{\pm}] = \pm J_{\pm}, \quad [J_+,J_-] = 2J_0\, .
\end{equation}
Using these operators, we construct the standard basis for representations labeled by $j = 0, \frac{1}{2}, 1, \frac{3}{2}, \cdots$. The basis states for representation $j$ are
\begin{equation}
    \mathcal{H}_j = \{ \ket{j,m} \mid m = -j, -j+1, \dots, j-1, j \}\, ,
\end{equation}
where $\CH_j$ is the Hilbert space of the irreducible representation. The generators act on these states as
\begin{equation}
    \begin{split}
        J_0 \ket{j,m} &= m \ket{j,m}\, , \\
        J_{\pm} \ket{j,m} &= \sqrt{j(j+1) - m(m\pm 1)} \ket{j, m \pm 1}\, .
    \end{split}
\end{equation}
We consider an initial state $\ket{\psi_0} \in \mathcal{H}_j$ and compute its Krylov complexity with respect to the $SU(2)$ generators. The time-evolved state is
\begin{equation}\label{psithetha}
    \ket{\psi (\theta_x, \theta_y, \theta_z)} = \exp\left( -i \vec{\theta}\cdot\vec{J}\right) \ket{\psi_0}\, ,
\end{equation}
where $\vec{J} = (J_x, J_y, J_z)$ and $\vec{\theta} = (\theta_x, \theta_y, \theta_z)$.
Choosing the initial state as the highest weight state
\begin{equation}
    \ket{\psi_0} = \ket{j,j}\, ,
\end{equation}
and assuming equal weights for all generators ($\mu_x = \mu_y = \mu_z = 1$), we proceed with the Gram-Schmidt orthogonalization by successively applying the algebra generators as described in Section~\ref{GKCdefinition}. This process leads to the following block structure
\begin{equation}
    \begin{split}
    &\mathcal{K}_0 = \text{span}\{ \ket{K_{0,0}}= \ket{j,j}\}, \qquad ~~~~~d_{\CK_0} = 1 
    \\
     &\mathcal{K}_1 = \text{span}\{ \ket{K_{1,0}}= \ket{j,j-1}\}, \qquad d_{\CK_1} = 1 
\\
&~~~~\vdots \\
& \mathcal{K}_i = \text{span}\{ \ket{K_{i,0}}= \ket{j,j-i}\},~ \qquad d_{\CK_i} = 1  
 \\
&~~~~\vdots \\
& \mathcal{K}_{2j} = \text{span}\{ \ket{K_{2j,0}}= \ket{j,-j}\},~ \qquad d_{\CK_{2j}} = 1  
\\
&\mathcal{K}_{{\CK_r}}=0, ~~~~~~~~~~~~~ r> 2j
    \end{split}
\end{equation}
The entire Krylov space is specified as 
\begin{equation}
    \mathcal{K} = \oplus_{i=0}^{2j} \mathcal{K}_i\, ,
\end{equation}
where $ \dim ( \mathcal{K})  =  \dim(\mathcal{H}_j)= 2j+1$, and thus, $ \mathcal{K} = \mathcal{H}_j$. This implies that the initial highest-weight state $\ket{\psi_0} = \ket{j,j}$ can evolve under the $\mathrm{SU}(2)$ algebra to reach any state in the Hilbert space $\mathcal{H}_j$. 
The Krylov complexity of the state \eqref{psithetha} can be found as the expectation value of the Krylov operator 
\begin{equation}
    \hat{K}_{\ket{\psi_0}} = \sum_{k=0}^{2j} k~ P_k = \sum_{k=0}^{2j} k~ \ket{ j, j-k}\bra{j,j-k}\, ,
\end{equation}
which yields
\begin{equation}
    C_\psi(\theta_x,\theta_y,\theta_z)= \sum_{k=0}^{2j} k~ |\bra{j,j-k}\exp \big( -i \vec{\theta}\cdot\vec{J}\big) \ket{j,j} |^2\, .
\end{equation}
To compute the complexity, we must evaluate matrix elements of the form
\begin{equation}\label{overlap}
    \bra{j,m} e^{-i \vec{\theta}\cdot\vec{J}} \ket{j,j},
\end{equation}
where $\vec{J} = (J_x, J_y, J_z)$ and $\vec{\theta} = (\theta_x, \theta_y, \theta_z)$.

The operator $e^{-i \vec{\theta}\cdot\vec{J}}$ represents a rotation by angle $|\theta| = \sqrt{\theta_x^2 + \theta_y^2 + \theta_z^2}$ about the axis $\hat{n} = \vec{\theta}/|\theta|$. Using the Euler angle parametrization \cite{varshalovich1988quantum}
\begin{equation}
    U(\vec{\theta}) = \exp \big( -i \vec{\theta}\cdot\vec{J}\big) = U(\alpha,\beta,\gamma) = \exp \big(-i \alpha J_z\big)\exp \big(-i \beta J_y\big) \exp \big(-i \gamma J_z\big)\, ,
\end{equation}
where we have 
\begin{equation}\label{Euler}
    \begin{split}
        \alpha= & ~\tan ^{-1}\Big(\frac{\theta_y}{\theta_x}\Big)+ \tan ^{-1} \Big(\frac{\theta_z}{|\theta|}\tan\big(\frac{|\theta|}{2}\big)\Big)- \frac{\pi}{2}\, ,
        \\
        \beta=&~ 2 ~\sin ^{-1}\Big( \sin\big(\frac{|\theta|}{2}\big) \sqrt{\frac{\theta_x^2+\theta_y^2}{|\theta|^2}}\Big)\, ,
        \\
        \gamma=& -\tan ^{-1}\Big(\frac{\theta_y}{\theta_x}\Big)+ \tan ^{-1} \Big(\frac{\theta_z}{|\theta|}\tan\big(\frac{|\theta|}{2}\big)\Big)+ \frac{\pi}{2}\, .
    \end{split}
\end{equation}
One can simplify the expression \eqref{overlap} in terms of Wigner D-matrix elements as \cite{wigner1931gruppentheorie}
\begin{equation}
   \bra{j,m}\exp \big( -i \vec{\theta}.\vec{J}\big) \ket{j,m'} \equiv D^j _{mm'}(\alpha,\beta,\gamma)\, ,
\end{equation}
where
\begin{equation}
    D^j _{mm'}(\alpha,\beta,\gamma) =\bra{j,m} \exp \big(-i \alpha J_z\big)\exp \big(-i \beta J_y\big) \exp \big(-i \gamma J_z\big) \ket{j,m'} = e^{-im\alpha} d^j_{mm'}(\beta)
    ~e^{-i\gamma m'}\, ,
\end{equation}
and
\begin{multline}
d^j_{mm'}(\beta) = \bra{j,m}  \big(-i \beta J_y\big) \ket{j,m'}=\\\sqrt{(j+m)!(j-m)!(j+m')!(j-m')!} \sum _{s_{min}}^{s_{max}} \frac{(-1)^{m-m'+s} \big(\cos\frac{\beta}{2}\big)^{2j+m'-m-2s}\big(\sin\frac{\beta}{2}\big)^{m-m'+2s}}{(j+m'-s)!~s!~ (m-m'+s)!~ (j-m-s)!}\, .
\end{multline}
This is the Wigner small $d$-matrix and $ s_{min}= \max (0,m'-m)$ , $ s_{max} = \min (j+m',j-m)$. 

\begin{figure}[h!]
    \centering
    \begin{subfigure}[b]{0.43\textwidth}
        \centering
        \includegraphics[width=\linewidth]{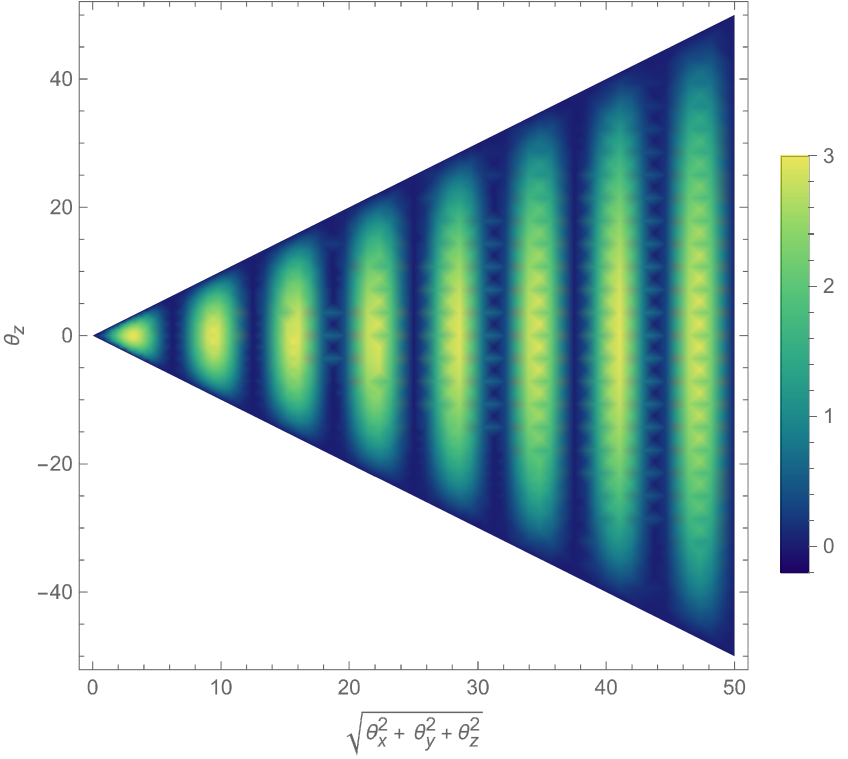}
        \caption{}
        \label{fig:subfig1}
    \end{subfigure}
    \hfill
    \begin{subfigure}[b]{0.43\textwidth}
        \centering
        \includegraphics[width=\linewidth]{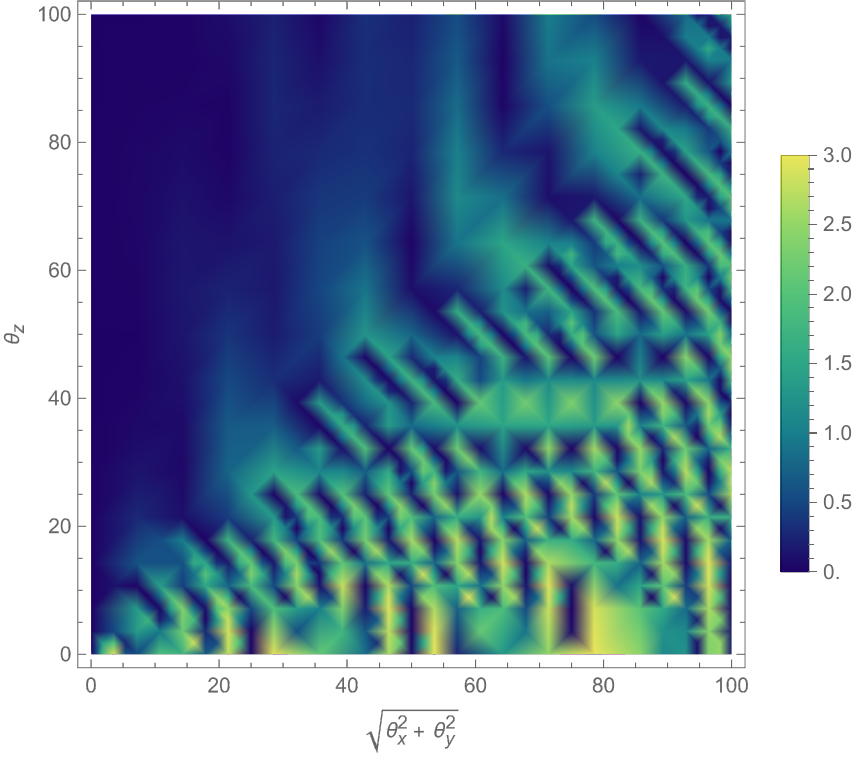}
        \caption{}
        \label{fig:subfig2}
    \end{subfigure}
    
    
    
    \caption{The complexity of the initial state $ \ket{j=3/2,m=3/2}$ under $SU(2)$ evolution.}
    \label{fig3}
\end{figure}

As a result, we have 
\begin{multline}
    \bra{j,j-k}\exp \big( -i \vec{\theta}.\vec{J}\big) \ket{j,j}= D^j_{j-k,j}(\alpha,\beta,\gamma)\\ = e^{-i \alpha (j-k)} ~d^j_{j-k,j}(\beta)~ e^{-i\gamma j} = e^{-i \alpha (j-k)} ~ e^{-i\gamma j}~ \sqrt{\frac{(2j)!}{(2j-k)!k!}} \Big(\cos \frac{\beta}{2}\Big)^{2j-k}~\Big(\sin \frac{\beta}{2}\Big)^{k}\, ,
\end{multline}
and thus, the generalized Krylov complexity of the state \eqref{psithetha} reads 
\begin{equation}
\begin{split}
        C_\psi(\theta_x, \theta_y,\theta_z)& = \sum _{k=1}^{2j} k \frac{(2j)!}{k!(2j-k)!} \Big(\cos ^2 \frac{\beta}{2}\Big)^{2j-k}~\Big(\sin^2 \frac{\beta}{2}\Big)^{k}
        \\
        & = 2j ~ \sin^2 \frac{\beta}{2}
        \\
        &= 2j~ \sin^2 \Big(\frac{|\theta|}{2} \Big) \frac{\theta_x^2+\theta_y^2}{|\theta|^2} = 2j~ \sin^2 \Big(\frac{|\theta|}{2} \Big) \frac{|\theta|^2 - \theta_z^2}{|\theta|^2}\, .
\end{split}
\end{equation}
Figure \eqref{fig3} displays the generalized Krylov complexity for the initial state with total angular momentum $j=3/2$ under $SU(2)$ algebra evolution.
\\

\begin{figure}[h!]
    \centering
    \begin{subfigure}[b]{0.43\textwidth}
        \centering
        \includegraphics[width=\linewidth]{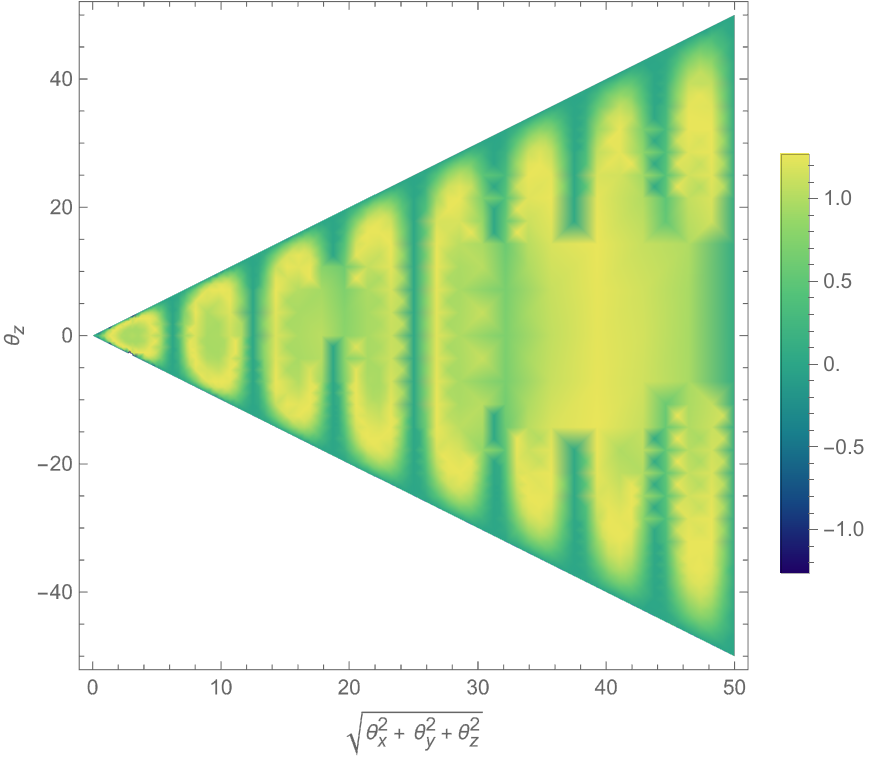}
        \caption{}
        \label{fig:subfig1}
    \end{subfigure}
    \hfill
    \begin{subfigure}[b]{0.43\textwidth}
        \centering
        \includegraphics[width=\linewidth]{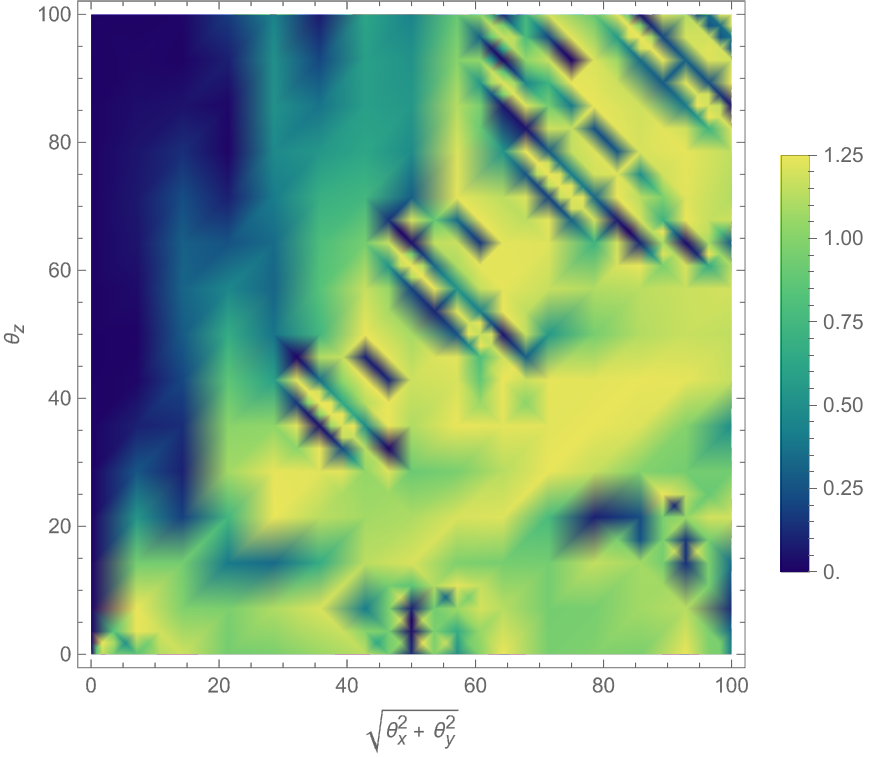}
        \caption{}
        \label{fig:subfig2}
    \end{subfigure}
    
    
    
    \caption{The complexity of the initial state $ \ket{j=3/2,m=1/2}$ under $SU(2)$ evolution.}
    \label{fig4}
\end{figure}

Let us now examine the case where the initial state is an arbitrary angular momentum eigenstate
\begin{equation}
    \ket{\psi_0} = \ket{j,m}\,.
\end{equation}
Applying the procedure defined in Section~\ref{GKCdefinition} to this state yields the following block structure
\begin{equation}
    \begin{split}
    &\mathcal{K}_0 = \text{span}\{ \ket{K_{0,0}}= \ket{j,m}\}, \qquad ~~~~~~~~~~~~~~~~~~~~~~~~~~~~~~~~~~~~~~~~~~~d_{\CK_0} = 1 
    \\
     &\mathcal{K}_1 = \text{span}\{ \ket{K_{1,0}}= \ket{j,m+1}, ~\ket{K_{1,1}}= \ket{j,m- 1} \}, ~~~~~~~~~~~\qquad d_{\CK_1} = 2
        \\
     &\mathcal{K}_2 = \text{span}\{ \ket{K_{2,0}}= \ket{j,m+2}, ~\ket{K_{2,1}}= \ket{j,m- 2} \}, ~~~~~~~~~~~\qquad d_{\CK_2} = 2
    \\
     &~~~~\vdots \\
     & \mathcal{K}_{\tilde{m}} = \text{span}\{ \ket{K_{\tilde{m},0}}= \ket{j,m+\tilde{m}},~ \ket{K_{\tilde{m},1}}= \ket{j,m-\tilde{m}}\}, \qquad~~~~~ d_{\CK_{\tilde{m}}} = 2  
     \\
     & \mathcal{K}_{\tilde{m}+1} =\left\{ \begin{array}{l}
       \text{span}\{ \ket{K_{\tilde{m}+1,0}}= \ket{j,m-\tilde{m}-1},~~~~~~~~\text{if}~~  m>0
       \\
        \text{span}\{ \ket{K_{\tilde{m}+1,0}}= \ket{j,m+\tilde{m} +1},~~~~~~~~\text{if}~~  m<0
       \end{array}\right.~~~ d_{\CK_{\tilde{m}+1}} =1
       \\ 
      &~~~~\vdots
      \\
          & \mathcal{K}_{j+|m|} =\left\{ 
    \begin{array}{l}
       \text{span}\{ \ket{K_{j+|m|,0}}= \ket{j,-j},~~~~~~~~\text{if}~~  m>0
       \\
        \text{span}\{ \ket{K_{j+|m|,0}}= \ket{j,+j},~~~~~~~~\text{if}~~  m<0
       \end{array}\right.~~ ~~~~~~~d_{\CK_{j+|m|}} =1
          \\
     & \mathcal{K}_{r}=0, ~~~~~~~~~~~~~ r> (j+|m|)
    \end{split}
\end{equation}
where $ \tilde{m} = j- |m|$, and the entire Krylov space is 
\begin{equation}
    \mathcal{K} = \oplus_{i=0}^{j-|m|} \mathcal{K}_i\, ,
\end{equation}
which again cover entire Hilbert space $\mathcal{H}_j$. The generalized Krylov operator, in this case, can be written as 
\begin{multline}
   \hat{K}_{\ket{\psi_)}} = \sum_{k=0}^{j+|m|} k~ P_k 
   \\= \sum_{k=0}^{j-|m|} k~ \Big ( \ket{j,m+k}\bra{j,m+k} +  \ket{j,m-k}\bra{j,m-k}\Big) + \sum_{k=j-|m|+1}^{j+|m|} k~  \ket{j,m-k}\bra{j,m-k}\, .  
\end{multline}
This holds for $ m>0$ while for $m<0$ the corresponding expression becomes
\begin{multline}
 \hat{K}_{\ket{\psi_0}} = \sum_{k=0}^{j+|m|} k~ P_k \\= \sum_{k=0}^{j-|m|} k~ \Big ( \ket{j,m+k}\bra{j,m+k} +  \ket{j,m-k}\bra{j,m-k}\Big) + \sum_{k=j-|m|+1}^{j+|m|} k~  \ket{j,m+k}\bra{j,m+k}\, .
\end{multline}
Therefore, putting everything together, we obtain the final expression for the generalised Krylov complexity as follows
\begin{equation}\label{jm}
    C_\psi (\theta_x,\theta_y,\theta_z) = \left\{ \begin{array}{l}
     \sum\limits{_{k=1}^{j+|m|}}~ k~ |d^j_{m-k,m}(\beta)|^2+ \sum_{k=1}^{{j-|m|}} ~ k~ |d^j_{m+k,m}(\beta)|^2,~~~~~~~~~~\text{if}~~ m>0
     \\
     ~~
     \\
     \sum_{k=1}^{{j+|m|}} ~ k~ |d^j_{m+k,m}(\beta)|^2 + \sum_{k=1}^{{j-|m|}} ~ k~ |d^j_{m-k,m}(\beta)|^2,~~~~~~~~~~\text{if}~~ m<0
     \end{array}\right.
\end{equation}
where $\beta$ can be written in terms of $\theta$ angles from \eqref{Euler}.
Figure \eqref{fig4} displays the Krylov complexity of the initial state for $ \ket{j=3/2,m=1/2}$ under $SU(2)$ evolution.
\\

\end{document}